# Coordination Engineering of Cu-Zn-Sn-S Aqueous Precursor for Efficient Kesterite Solar Cells

Linbao Guo[1,3,†], Jiangjian Shi[1,†], Qing Yu[1,3], Biwen Duan[1,3], Xiao Xu[1,3], Jiazheng Zhou[1,3], Jionghua Wu[1,3], Yusheng Li[1,3], Dongmei Li[1,3,4], Huijue Wu[1], Yanhong Luo[1,3,4,*] and Qingbo Meng[1,2,4,*]

1 Beijing National Laboratory for Condensed Matter Physics, Institute of Physics, Chinese Academy of Sciences, Beijing 100190, China

2 Center of Materials Science and Optoelectronics Engineering, University of Chinese Academy of Sciences, Beijing 100049, China

3 School of Physical Sciences, University of Chinese Academy of Sciences, Beijing 100049, China

4 Songshan Lake Materials Laboratory, Dongguan, Guangdong 523808, China

† These authors contributed equally to this work: Linbao Guo, Jiangjian Shi.

* E-mail: yhluo@iphy.ac.cn, qbmeng@iphy.ac.cn

**Abstract:** Aqueous precursors provide an alluring approach for low-cost and environmentally friendly production of earth-abundant $Cu_2ZnSn(S,Se)_4$ (CZTSSe) solar cells. The key is to find an appropriate molecular agent to prepare a stable solution and optimize the coordination structure to facilitate the subsequent crystallization process. Herein, we introduce thioglycolic acid, which possesses strong coordination (-SH) and hydrophilic (-COOH) groups, as the agent and use deprotonation to regulate the coordination competition within the aqueous solution. Ultimately, metal cations are adequately coordinated with thiolate anions, and carboxylate anions are released to become hydrated to form an ultrastable aqueous solution. These factors have contributed to achieving CZTSSe solar cells with efficiency of as high as 12.2% (a certified efficiency of 12.0%) and providing an extremely wide time window for precursor storage and usage. This work represents significant progress in the non-toxic solution fabrication of CZTSSe solar cells and holds great potential for the development of CZTSSe and other metal sulfide solar cells.

Photovoltaics have made great contributions to the release of global energy and environmental issues.[1] Kesterite $Cu_2ZnSn(S,Se)_4$ (CZTSSe) is one of the most environmentally friendly and inexpensive semiconductor light-absorbing materials for photovoltaic applications because of its non-toxic and earth-abundant components.[2-4] CZTSSe exhibits high light absorption of $>10^4$ cm$^{-1}$, an adjustable bandgap matching the solar spectrum,[5-9] high thermodynamic and environmental stability[10-12] and a device manufacturing process compatible with current thin film solar cells.[13-18] Solution processing of the CZTSSe thin film deposition by intermixing of precursor components at the molecular level has advantages of composition uniformity and morphology control over the conventional vacuum technique.[19-22] The hydrazine solution approach, with the advantages of high reduction and coordination ability,[23, 24] has resulted in the most efficient CZTSSe solar cells.[25] These positive results have encouraged scientists to explore the green solvent technique for CZTSSe fabrication, an ultimate trend toward non-vacuum semiconductor device production,[26-28] which has led to the development of a variety of solvent systems.[29-32] Certainly, among these systems, the aqueous system is the most alluring candidate from the perspective of safety, environmental effects and cost.

As early attempts at aqueous precursor systems, metal salt precursor routes such as chemical bath deposition (CBD),[33] successive ionic layer adsorption and reaction (SILAR)[34, 35] and electrochemical deposition have been explored.[36, 37] Spin coating using a metal salt-thiourea solution has yielded moderate efficiencies.[38, 39] Strategies for synthesizing nanocrystals from aqueous solutions or preparing colloid dispersions have also been developed. Cell performance has been obviously improved through pre-synthesis of the colloid dispersion and detailed adjustment of the element composition.[40, 41] However, the resulting device efficiency is still much lower than that of the organic systems used to date.[42] This may arise from residual anions such as chloride ions and long-chain stabilizers, which have a significant impact on CZTSSe crystallization.[43-46] In addition, these aqueous precursor solutions usually suffer from low stability since the formed metal-thiourea complexes and the dispersed colloids easily precipitate due to the lack of strong solute-solvent interactions.[47] Therefore, more judicious engineering of the CZTSSe aqueous precursors is desired to increase the current device performance.

Generally, metal sulfide compounds or complexes are insoluble or easily precipitate within an aqueous solution due to poor hydration.[48] Thus, molecular engineering of organic ligands with hydrophilic groups may help stabilize CZTSSe (or Cu, Zn and Sn) aqueous precursors. Thiols engage in strong coordination interactions with metal cations and have been widely used in the chemical synthesis of metal sulfide films in non-aqueous precursor systems.[49-51] Thiols with hydrophilic groups, such as carboxyl, hydroxyl and amine moieties, should be useful in constructing CZTSSe aqueous precursors.[52] Thioglycolic acid (TGA), which has a lower GHS risk level and low cost, is the most promising water-soluble ligand.[53] Owing to these potential advantages, TGA was once used for aqueous CZTSSe precursor preparation.[54-56] However, the resulting efficiency of 7.38% is still much lower than that of the hydrazine and other organic systems. This mainly arises because (1) the fundamental coordination chemistry of this aqueous system has not been clearly understood and (2) that how to precisely manipulate the complexes structure of TGA-metal to facilitate its application in CZTSSe solar cells has not been studied. Herein, we systematically investigate the coordination engineering in the Cu-Zn-Sn-S aqueous precursor solution. The competition between metal-S and metal-COO coordination interactions has been manipulated by adjusting the deprotonation degree of TGA in solution. In the optimal structure, the metal cation is adequately coordinated with thiolate anions, and the carboxylate anions are released to become fully hydrated, thus facilitating the formation of an ultrastable metal-TGA aqueous solution and improving the nucleation and crystallization of the CZTSSe thin film. Our strategies enable the fabrication of CZTSSe solar cells with a remarkable efficiency of 12.2% and a certified efficiency of 12.0%, which is the best result for CZTSSe solar cells among the aqueous precursor systems investigated to date. This significant progress in the non-toxic solution fabrication of CZTSSe solar cells brings great promise for the future development of CZTSSe solar cells.

**Results and Discussion**

**Coordination competition between metal-S and metal-O**

As a preliminary attempt, metal oxides, such as CuO, ZnO and SnO, were added to an aqueous TGA solution, and the mixture was continuously stirred for several days.

Unfortunately, these metal oxides have limited solubility in the TGA solution (Supplementary Figure 1) and the obtained mixtures cannot be used as precursors for CZTSSe film deposition. Nonetheless, we found chemical reactions had occurred within these solid-liquid mixtures, and new complexes formed, as evidenced by the apparent colour change of the added solids. For instance, the colour of the CuO powder changed from black to white, that of SnO changed from black to light brown, and the precipitate in the ZnO system remained white but fluffy (Figure 1(a)). These results are positive because chemical or coordination reactions are usually the first step in the dissolution of metal oxides, sulfides or elementary substances.

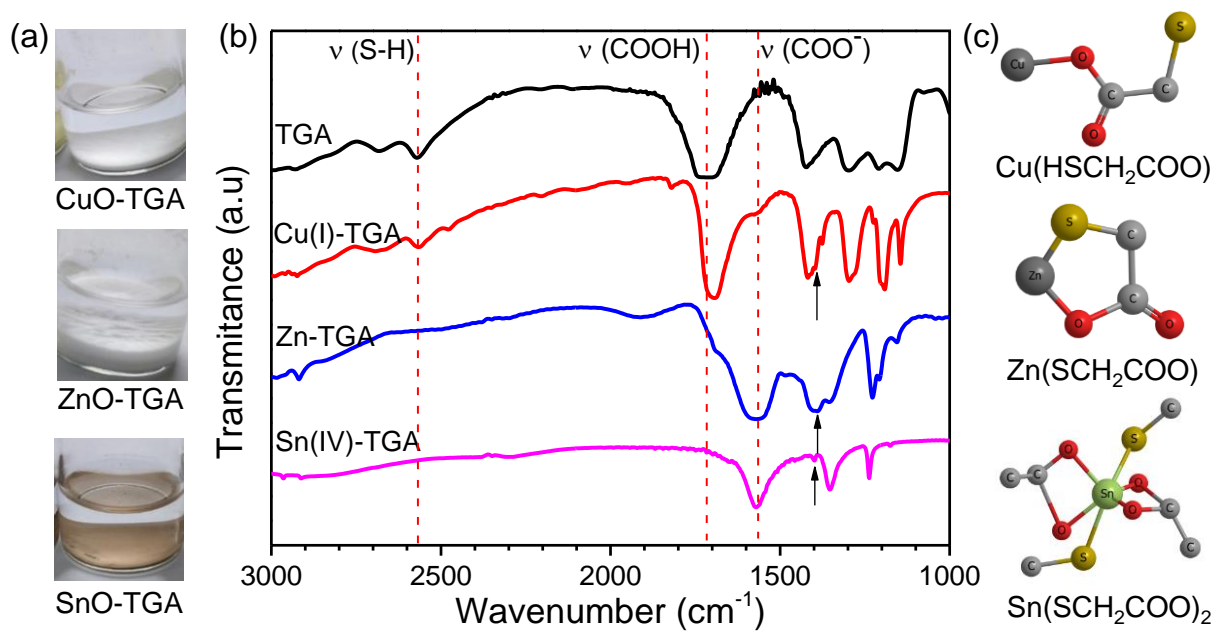

**Figure 1. Chemical reactions between metal oxides and TGA.** (a) Photographs of the metal oxide-TGA reaction products in water. (b) Fourier transform infrared spectra of the collected products and pure TGA. Characteristic peaks and their corresponding vibration modes that can reflect the reaction processes are marked by dashed lines and arrows. The change in the atomic valence of Cu or Sn arose from redox reactions with TGA or di-TGA. (c) Proposed structures of the metal oxide-TGA insoluble products. Within these structures, almost all the carboxylate groups are coordinated to metal atoms rather than exposed to the surrounding water molecules to facilitate dissolution.

Chemical reactions between these metal oxides and TGA in an aqueous solution were investigated by Fourier transform infrared (FTIR) spectroscopy of the precipitation products (Figure 1(b)). Notably, Cu(II) was reduced to Cu(I) in the presence of excess TGA, resulting in the formation of HOOCCH$_2$S·SCH$_2$COOH (di-TGA) as a byproduct.[57] The presence of di-TGA was confirmed according to $^{13}$C nuclear magnetic resonance (NMR) spectra of the supernatant (Supplementary Figure 2). We further found that Sn(II) can be oxidized to Sn(IV) by the di-TGA while generating TGA as coordinating ligand. When both Cu(II) and Sn(II) were present in the system, a redox reaction occurred between them to form Cu(I) and Sn(IV).[30] Thus to imitate the actual reaction within the Cu-Zn-Sn-S precursor in the spectra study of individual metal oxide-TGA systems, stoichiometric di-TGA was added to oxidize Sn(II) to Sn(IV). In the spectra of pure TGA, characteristic absorption bands for both the S-H and COOH groups are clearly observed at ~2570 cm$^{-1}$ and ~1714 cm$^{-1}$, respectively.[58-61] Changes in the absorption bands are observed after the reaction. First, the disappearance of the S-H stretching vibration in the spectra of the ZnO-TGA and SnO-TGA products implies that the S-H group was deprotonated, allowing S to coordinate to the metal atom.[58] This change was realized through a metal oxide-thiol reaction, releasing H$_2$O as another product. Second, a new peak at ~1570 cm$^{-1}$ is observed for all three solid products, which mainly arises from the asymmetric stretching of COO$^-$ and is accompanied by a symmetric stretching peak in the low-frequency region,[58, 61] as depicted by arrows in Figure 1(b). This FTIR spectral change indicates that the COOH group was also deprotonated through the metal oxide-acid reaction, resulting in coordination between the oxygen (COO$^-$ group) and the metal atoms. Unlike the reaction products of ZnO and SnO with TGA, the products of the reaction with CuO still gave rise to a S-H absorption band, most likely because Cu coordinated only with COO$^-$ to form HSCH$_2$COOCu at low pH.[62] On the basis of these spectral signatures, possible reactions between TGA and these metal oxides were proposed (Supplementary Scheme 1), and possible structures of the produced metal-TGA complexes are presented in Figure 1(c). It is clear that both deprotonated thiol and carboxyl groups coordinated to Sn and Zn, while only the carboxylate group coordinated to Cu. In these structures, the polarity of the carboxylate group is significantly weakened, and its hydrophilic characteristics cannot be fully expressed. Therefore, to enhance the solubility of the metal-TGA complexes, the carboxylate group must be adequately released to be hydrated by the surrounding water molecules by substituting the metal-carboxylate coordination with metal-thiolate coordination.

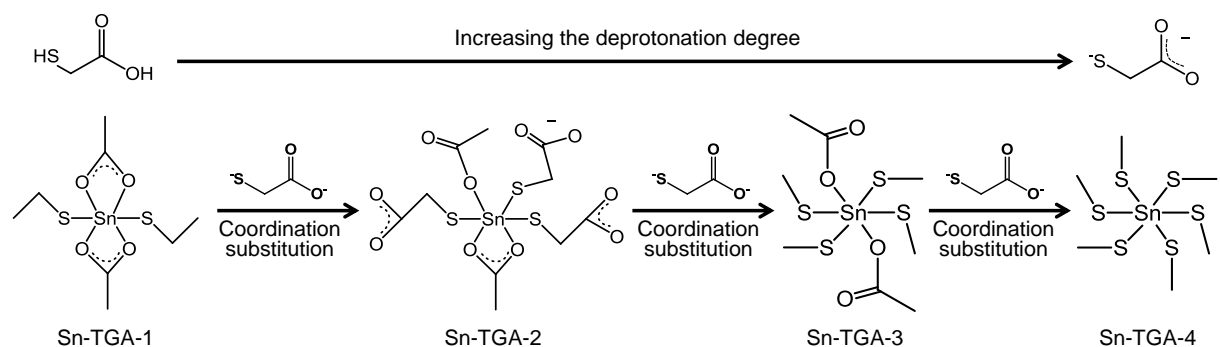

**Figure 2. Coordination engineering route within the aqueous Sn-TGA solution.** Deprotonation of the thiol group could enhance its coordination to the Sn atom and thus facilitate substitution of the Sn-COO coordination. Different degrees of deprotonation will result in varied Sn-TGA complexes. Ultimately, the Sn is completely coordinated with the S$^-$ groups, and the COO$^-$ is fully exposed to hydrate with the solvent surroundings, which helps to form a stable Sn-TGA aqueous solution.

**Engineering of coordination structures**

However, within the acidic or neutral TGA/water environment, metal-thiolate coordination cannot easily be fully realized because thiol groups are usually difficult to deprotonate to form R-S$^-$ (R: HOOCCH$_2$) with strong metal complexing ability.[63-66] Thus, the above desired substitution process can be achieved only when ionization (i.e., deprotonation) of the thiol group of the TGA is enhanced.[49] Figure 2 schematically presents the O donor and S donor coordination substitution processes that may occur with increasing TGA deprotonation in the aqueous metal oxide-TGA solution, where Sn compounds are used as an example. For the initial state (Sn-TGA-1), a double decomposition reaction occurs between SnO and TGA to form salt and water, and both oxygen atoms of the carboxylate group are coordinated to the Sn atom, forming a bidentate carboxylate structure. When the deprotonation degree of TGA in solution is increased, the highly concentrated and activated R-S$^-$ can substitute for one of the Sn-O coordination bonds of the bidentate carboxylate. This will introduce an additional Sn-S bond and a free carboxylate terminal group and thus help enhance the hydrophilicity of the Sn-TGA complex. Upon further deprotonation of the thiol group, more Sn-O coordination of the bidentate carboxylate can be substituted by Sn-S bonds, forming the Sn-TGA-3 structure and ultimately the Sn-TGA-4 structure. Within Sn-TGA-4, the Sn atom is completely coordinated by R-S$^-$, and all the carboxylate groups are exposed to water molecule surroundings. This coordination structure should be able to form a more stable (or the most stable) Sn-TGA aqueous solution. For the Cu-TGA and Zn-TGA complexes, similar coordination substitution processes should also be realized (Supplementary Figure 3), thus helping to prepare stable and effective CZTSSe precursors.

The above coordination substitution processes depend highly on deprotonation of the -SH group. Increasing the pH of the solution is the simplest, most reliable and most widely used method to facilitate this process.[67, 68] Ammonia (NH$_3$·H$_2$O) is the best candidate for this purpose because it does not introduce any metal or carbon impurities and can be easily removed at relatively low temperatures. These advantages prompted us to use ammonia to regulate the aqueous system, and positive results were obtained. The molar ratio between ammonia and TGA was the main variable, while the concentrations of both TGA (4 M) and the metal (0.4 M) were held constant for simplicity. Distinct variations in the Cu-Zn-Sn-TGA system were observed when the amount of ammonia was increased (Supplementary Table 1 and Supplementary Figure 4). Specifically, the metal oxides were completely dissolved when the NH$_3$/TGA ratio approached 1.0, and the solution colour changed from dark amber to light yellow when the ratio approached 2.0 (Supplementary Figure 5). These phenomena indicate changes in the metal-TGA complexing structures within the aqueous solution. Solution Raman scattering and NMR spectroscopies, which are sensitive to coordination structure and chemical environment, were used to trace the coordination chemistry of this solution system.[51, 69] To obtain comprehensive information, Cu-Zn-Sn-TGA, Cu-TGA, Zn-TGA and Sn-TGA were all studied in our measurements.

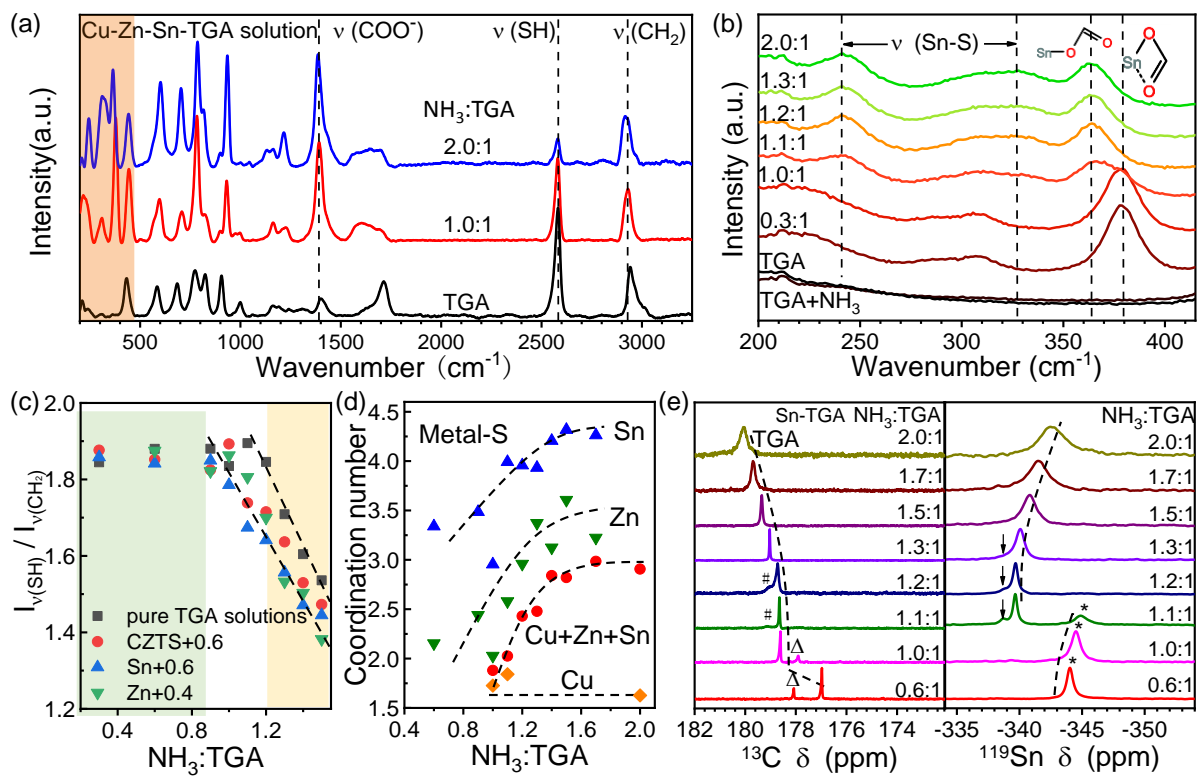

**Figure 3. Raman scattering and NMR characterizations of coordination structure variations.** (a) Raman spectra of pure TGA and the Cu-Zn-Sn-TGA aqueous solution with varied NH$_3$/TGA ratios. Signals corresponding to stretching ($v$) of COO$^-$, SH and CH$_2$ groups are marked with dashed lines. (b) Amplified low-wavenumber Raman spectra, illustrating Sn-O and Sn-S vibrations and their variations with the NH$_3$/TGA ratio. (c) Intensity ratio of the $v$ (SH) ($I_{v(SH)}$) and $v$ (CH$_2$) ($I_{v(CH2)}$) Raman signals at various NH$_3$/TGA ratios. The intensity of $v$ (CH$_2$) ($I_{v(CH2)}$) is used as an internal standard here. For the pure TGA solution, the SH group remains in a protonated state until the ratio is higher than 1.2. (d) Approximate calculation of the metal-S coordination number according to the Raman intensity difference between the TGA solution and the metal-containing solution. The metal-S coordination number increases obviously with the NH$_3$/TGA ratio. (e) $^{13}$C NMR spectra of the COOH group and $^{119}$Sn NMR spectra to confirm variations in the coordination structure of Sn-TGA. The non-solvent NMR signals are marked with 'Δ' or '#'. Dashed lines are visual guides for changes inducing increasing by NH$_3$/TGA ratio.

Within the wide-region Raman spectra, the characteristic scattering signals attributable to SH

and CH$_2$ stretching at ~2580 cm$^{-1}$ and ~2935 cm$^{-1}$, respectively, can be seen for all aqueous solutions (Supplementary Figure 6-7).[67, 68] The C=O stretching frequency in the pure TGA aqueous solution is 1714 cm$^{-1}$. The apparently depolarized CH$_2$ bending frequency at 1395-1400 cm$^{-1}$, clearly visible in the un-ionized acid, is present in all states of deprotonation. Upon ionization of the carboxyl group, the Raman peak at 1395-1400 cm$^{-1}$ is partially masked by the very intense symmetrical stretching frequency of the ionized carboxyl, and the asymmetrical stretching frequency of the latter (1550-1575 cm$^{-1}$) also appears. The $v$ (CH$_2$) intensity at ~2935 cm$^{-1}$ is almost independent of the amount of ammonia, which is in good agreement with the unchanged TGA concentration and thus confirms that other optical properties within the Raman process, such as the light penetration depth and other scattering processes (e.g., Mie scattering and Rayleigh scattering), have not been influenced by the solution conditions (Supplementary Figure 7-8). In the pure TGA solution, under a low NH$_3$/TGA ratio, the $v$ (SH) intensity remained nearly unchanged, and the $v$ (COO$^-$) intensity increased gradually. At higher NH$_3$/TGA ratios, the $v$ (SH) intensity was significantly reduced. After addition of the metal oxide, the $v$ (SH) intensity decreased by a constant value compared with that of the TGA solution, indicating that deprotonation of the S-H group occurred and that R-S$^-$ coordinated with the metal. In addition, more abundant spectroscopic signatures arising from Sn atom-associated vibrations were observed in the low wavenumber region, amplified in Figure 3(b). At NH$_3$/TGA ratios lower than 1.0, strong scattering signals appeared at ~380 cm$^{-1}$. These signals abruptly vanished and a slightly lower-frequency vibration mode appeared at 365 cm$^{-1}$ when the NH$_3$/TGA ratio exceeded 1.2. At a ratio of 1.1, these two scattering signals coexisted within the spectrum. Both of these Raman signals can be attributed to the stretching vibration of the Sn-O bond, while the frequency difference implies modified coordination geometries,[67] as depicted in the inset of Figure 2(b). An O···Sn-O bidentate structure is expected to possess a tighter Sn-O Coulomb interaction and thus a higher bond stretching frequency. Accompanying the vanishing of this bidentate structure, new Raman scattering peaks were observed at 240 cm$^{-1}$ and 330 cm$^{-1}$. These two peaks were confirmed to arise from the stretching vibration of the Sn-S bond[71-73] by obtaining Raman spectra of the Sn-mercaptoethanol solution, which possesses only Sn-S coordination (Supplementary Figure 9). Therefore, based on the changes in the low-frequency Raman scattering signals, we can confirm that the coordination substitution reaction from Sn-O to Sn-S indeed occurred when the $NH_3$/TGA ratio exceeded 1.0. This finding agrees well with the substitution processes schematically presented in Figure 2. For the coordination of other metal atoms, we believe similar processes can also occur, although they cannot be directly probed by Raman scattering.

This replacement of one of the bidentate carboxylate O atoms by $S^-$ consumes more $R-S^-$ groups and thus results in a reduction in the SH concentration (Supplementary Scheme 2), which can be quantitatively reflected by the $v$ (SH) intensity in the Raman spectra.[64] The $CH_2$ band at 2935 $cm^{-1}$ was chosen as an internal standard since all the evidence indicated that the intensity of this band was not substantially affected by the change in the $NH_3$/TGA ratio (Supplementary Figure 8). For clarity, the initial $v$ (SH) intensity relative to the $v$ ($CH_2$) intensity ($I_{v\,(SH)}/I_{v\,(CH2)}$) for these samples is set to be the same by adding certain values. It is apparent that in the pure TGA solution, the $I_{v\,(SH)}/I_{v\,(CH2)}$ value remains unchanged when the ratio is lower than 1.2 and then exhibits an obvious decline, indicating significant deprotonation of the SH group. When the metal oxides are dissolved, a certain proportion of SH is consumed in the initial stage to react with metal oxides and coordinate to the metal atoms. At low $NH_3$/TGA ratios of <0.9, this initial $I_{v\,(SH)}/I_{v\,(CH2)}$ value remains constant, implying an unchanged metal-TGA coordination structure. Compared to the solvent itself, this $I_{v\,(SH)}/I_{v\,(CH2)}$ value begins to decrease at an obviously lower $NH_3$/TGA ratio of ~0.9, implying that the coordination substitution (from metal-S to metal-O) accelerated the pH-induced deprotonation of the SH group. This result is strong support for the observed coordination substitution process. From these intensities, we can further estimate the coordination number of the metal-S (Supplementary Note 1) and trace the variations. Clearly, the coordination number of all the metal atoms has increased at higher $NH_3$/TGA ratios (Figure 3d), further confirming the coordination substitution process that we have proposed.

More evidence was gathered from independent NMR characterizations (Figure 3(e)). Within the Sn-TGA solution, two types of carboxyl $^{13}C$ chemical shifts were observed at a low $NH_3$/TGA ratio. The chemical shift at δ = 177 ppm (marked by dashed lines) was attributed to the free COOH or $COO^-$ within the solvent, while another downfield chemical shift at δ=178 ppm (triangles) was attributed to the carboxyl carbon of the Sn-OOC because the donor coordination weakened the shielding effect. Free carboxyl groups existed in all studied NH$_3$/TGA ratios, while the Sn-OOC NMR signal disappeared when the ratio exceeded 1.0. A new weak signal appeared at δ = 178.6 ppm (# marks) at a ratio of 1.1 but disappeared at ratios higher than 1.2. This signal was also ascribed to the Sn-OOC but with a modified coordination geometry.[74] The complete disappearance of the Sn-OOC coordination signal indicates that the proportion of Sn-O bonds among the Sn-TGA coordination was too low to be detected when the ratio was higher than 1.3. The $^{119}$Sn NMR signal from the bidentate carboxylate Sn-O bonding was initially observed at δ = -344 ppm (asterisks) and shifted upfield as the NH$_3$/TGA ratio increased. When the ratio exceeded 1.1, this signal completely disappeared, and a new NMR signal appeared at a chemical shift of δ = -339.5 ppm. This large downfield shift also implies a new Sn-TGA coordination structure, which strongly supports the appearance of the added Sn-S bonds. The Sn-S coordination possesses a long bond length and thus has a weak shielding effect on the Sn nucleus. Accompanying the Sn-S bond signal, a shoulder signal also appeared on the downfield side and disappeared at high NH$_3$/TGA ratios. Finally, broadened and upfield-shifted Sn-S signals dominated the NMR spectra, demonstrating that the Sn-TGA had a stable coordination structure with a high density of covalent electrons in the Sn-S bond and fast molecule exchange with the free TGA in the solution. This fast exchange behaviour indicates that the metal-TGA coordination system was adequately assimilated into the water molecule surroundings, which should be helpful for solution stability. Overall, we have realized the dissolution of the Cu-Zn-Sn-TGA system into an aqueous solution through thiol deprotonation, and the coordination structure can be adjusted by the NH$_3$/TGA ratio.

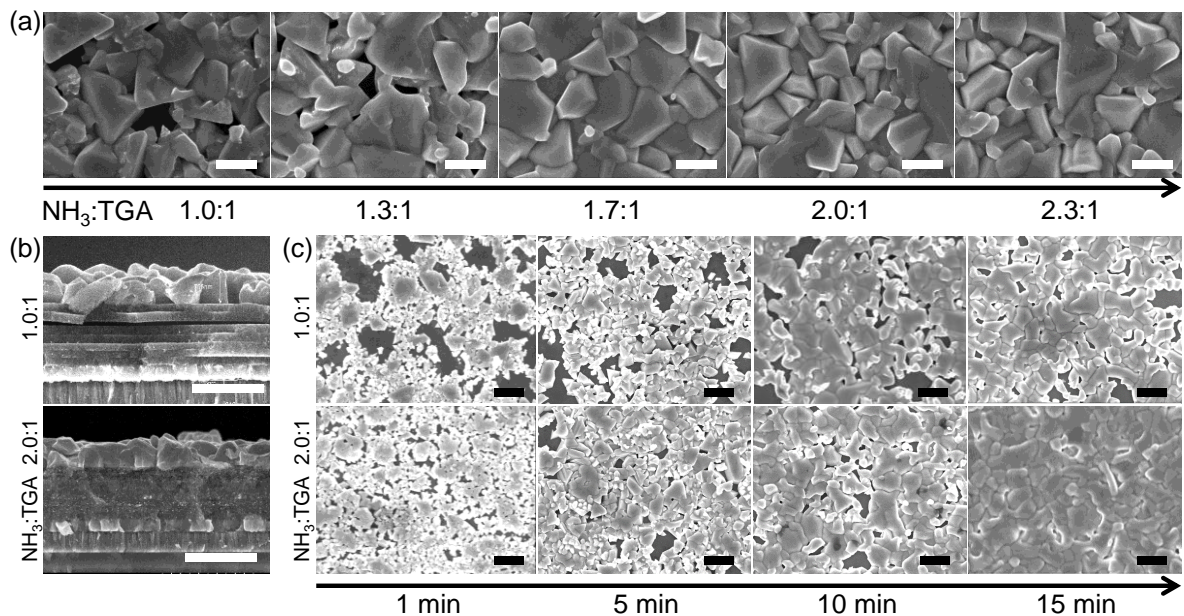

**Figure 4. Morphology of the CZTSSe films.** (a) Top-view and (b) cross-sectional scanning electron microscopy (SEM) images of the CZTSSe films derived from precursors possessing varied NH$_3$/TGA ratios. Selenization temperature: 520 °C; selenization duration: 15 min. Scale bar: 1 μm. (c) Selenization duration-dependent morphology evolution of the CZTSSe films at NH$_3$/TGA ratios of 1.0:1 and 2.0:1, respectively. Scale bar: 2 μm.

**Thin film deposition and solar cell performance**

These fully soluble solutions with an NH$_3$/TGA ratio of ≥1.0 were used as precursors for thin film deposition (spin coating), selenization and final fabrication of CZTSSe solar cells. The NH$_3$/TGA ratio (i.e., Cu-Zn-Sn-TGA coordination structure) significantly affected the morphology of the CZTSSe film after selenization at 520 °C for 15 min. As in the top-view scanning electron microscopy (SEM) images, large voids were observed for the samples derived from the precursors with NH$_3$/TGA molar ratios of 1.0:1 and 1.3:1. A dense and void-free film surface was realized only when the NH$_3$/TGA molar ratio exceeded 1.7. These differences are further reflected by the cross-sectional SEM images. For the 1.0-ratio sample, an interlayer boundary due to the multiple spin coating and severe interlayer separation were clearly seen in the fine grain layer. This means that mass transport and exchange during the selenization process were not sufficient for this sample. Comparatively, the 2.0-ratio sample exhibited a dense-packed fine grain layer as well as a relatively smooth and large top grain layer. These morphological differences arise from the distinct nucleation properties between these two samples. The 2.0-ratio sample possessed a much higher nucleation velocity and density than the other samples, serving as an important foundation for subsequent crystallization and ripening processes.

Essentially, these nucleation and mass transport properties are closely correlated with the activity of the metal elements within the CZTSSe precursor film. For the sample with a low $NH_3$/TGA ratio, carboxylate groups occupied a significant proportion of coordination sites around the metal atoms. The strong binding between the bidentate oxygen and the metal atoms that remained in the precursor film limited the interdiffusion ability of the metal atoms and thus suppressed the nucleation and mass transport processes.[75] Comparatively, metal sulfides exhibited much higher reactivity within the selenization process, thus affording better crystallization morphology.

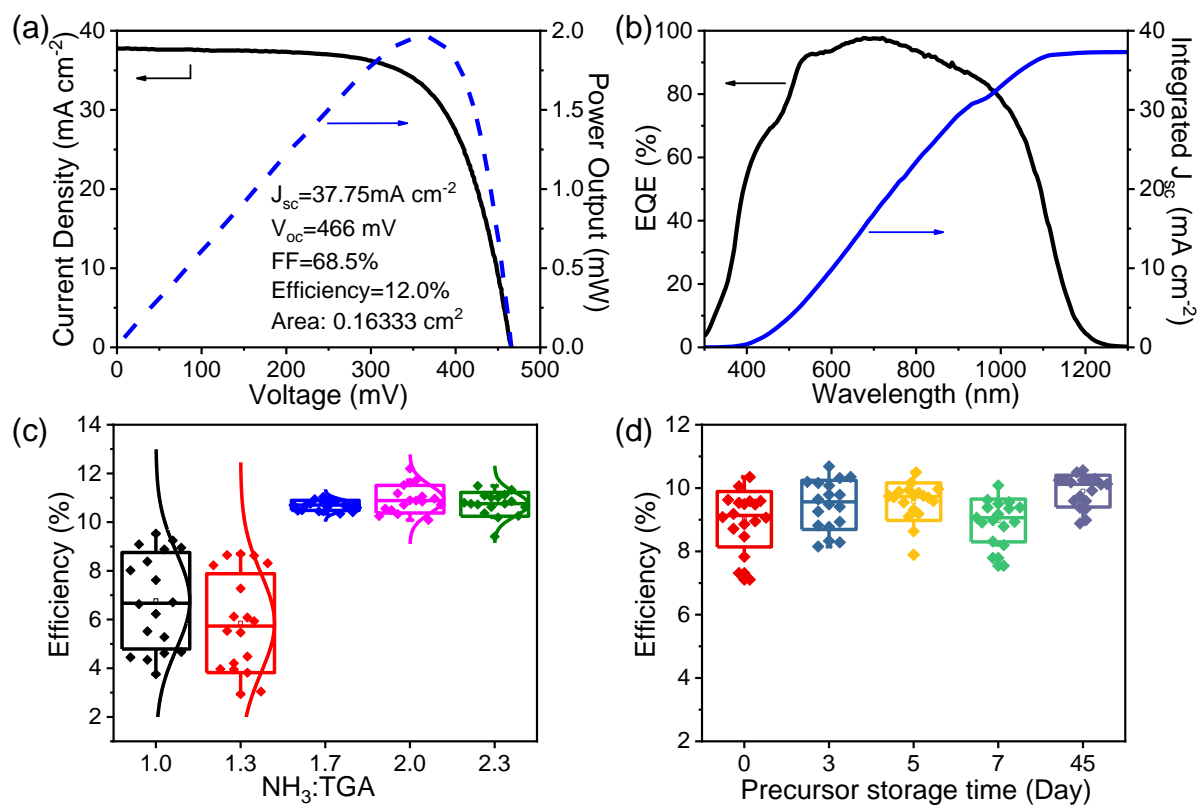

**Figure 5. Solar cell characterizations.** (a) Certified current-voltage characteristics of the CZTSSe solar cell derived from a precursor with an $NH_3$/TGA ratio of 2.0. The cell exhibits an efficiency of 12.0% under an effective area of 0.16333 cm$^2$. (b) External quantum

efficiency (EQE) spectrum of the certified cell, giving an integrated short-circuit current density of ~37 mA cm$^{-2}$. (c) NH$_3$/TGA ratio-dependent cell efficiency. (d) Cell efficiency as a function of precursor solution storage time. After storage for 45 days, the precursor solution can still be used to fabricate efficient solar cells.

CZTSSe solar cells were fabricated with a device configuration of glass/Mo/CZTSSe/CdS/ZnO/ITO/Ni/Al. The optimal cell exhibited a high efficiency of 12.2% in our laboratory measurement (Supplementary Figure 10), which is the highest among the aqueous precursor systems studied to date and is comparable to that of the DMSO system.[42] Certification of these CZTSSe cells yielded a high efficiency of 12.0% (Supplementary Figure 11) with an effective area of 0.16333 cm$^2$. Integration of the external quantum efficiency (EQE) of the cell resulted in a short-circuit current density ($J_{SC}$) of ~37 mA cm$^{-2}$, agreeing well with the current-voltage curve. The EQE spectrum indicates that the CZTSSe film possessed a bandgap of ~1.1 eV (Supplementary Figure 12). The cell exhibited an open-circuit voltage deficit of ~0.618 V, implying that we still have much room to improve the cell performance. The correlation between the cell efficiency and the NH$_3$/TGA ratio is presented in Figure 5(c). High efficiency can be achieved only when metal-S bonds dominate the coordination (NH$_3$/TGA ratio ≥ 1.7). This on the one hand arises from a better CZTSSe film morphology and, on the other hand, benefits from a lower material defect density and slower charge recombination velocity (Supplementary Figure 13). Owing to the stable metal-TGA coordination structure and its adequate hydration through the exposed carboxyl groups, the aqueous precursor solution exhibited ultrahigh stability for several months without the appearance of any solid precipitation or opaque suspensions. These stored precursors can still be used to fabricate efficient solar cells without any decline in efficiency, as presented in Figure 5(d). This result highlights another advantage of our system, that is, precursor stability, which should be considerably higher than those of previous systems based on nanocrystal or colloid dispersions. Such high precursor stability provides an extremely wide time window for the industrial production of CZTSSe solar cells.

**Conclusions**

The coordination structure of the metal-TGA has been systematically engineered by promoting deprotonation of the thiol group. Through engineering, the metal-S bond dominates the coordination, and the terminal carboxyl group is adequately exposed to hydration by the surrounding water to form an ultrastable precursor at the molecular level. This coordination structure affords better CZTSSe nucleation and crystallization and contributes to reduced electronic defects. These advantages contribute to achieving a remarkable efficiency up to 12.2% and a certified efficiency of 12.0%. In addition, this ultrastable precursor provides an extremely wide time window for CZTSSe solar cell fabrication. These results represent significant progress in the non-toxic solution fabrication of CZTSSe solar cells and provide great promise for the future development of CZTSSe and other metal sulfide solar cells.

## Experimental Methods

**Materials.** Copper (II) oxide (CuO, 99.9%) and selenium particle (Se, 99.999%) were purchased from Zhongnuo Advanced Material Technology Company. Tin(II) oxide (SnO, 99.9%), thioglycolic acid ($HSCH_2COOH$, 98%), cadmium sulfate ($CdSO_4$, 99%), thiourea ($NH_2CSNH_2$, 99%) and dithiodiglycolic acid (di-TGA) (($CH_2COOH)_2S_2$, 96%) were purchased from Aladdin Inc. Zinc (II) oxide (ZnO, 99.99%) was purchased from Sigma-Aldrich. Ammonia solution ($NH_3·H_2O$, 28%) were purchased from Beijing Chemical Industry Group CO., LTD. All chemicals were used as received without any further purification.

**Solution preparation.** All solutions were prepared at room temperature in a $N_2$-filled glovebox. The solubility limits were determined after overnight stirring. In all solutions for Raman and NMR characterization, the concentration of TGA was maintained as 4 M and specific amounts of $NH_3$ was added according to the $NH_3$/TGA ratio. The precursor solution for CZTS film deposition was obtained by dissolving 0.280 g CuO, 0.195 g ZnO and 0.269 g SnO in a mixture aqueous solutions of 4 M thioglycolic acid and varied $NH_3·H_2O$.

**Metal-TGA complex preparation.** Metal oxides (1 mmol) were mixed with 2.0 mL of TGA and 1.0 mL of water under magnetic stirring at 60 °C for 2 days. After sufficient reaction, the Metal-TGA complex precipitations were formed. The Metal-TGA complexes were separated by centrifuging and cleaned with water for 3 times. The Metal-TGA complexes were dried in vacuum box for 24 h at room temperature.

**Film deposition.** The precursor film of CZTSSe was deposited on Mo-coated soda lime glasses (0.3 Ω sq$^{-1}$, DC-magnetron sputtering deposited) by spin-coating the precursor solution at 4000 rpm for 20 s in a $N_2$-filled glove box. The as-deposited films were sintered on a hot plate at 400 °C for 2 min to eliminate organic residues. The spin-coating and annealing steps were repeated several times until the thickness of the precursor film reaches 1.0 μm. Finally, these thin films were selenized under Se/$N_2$ atmosphere at 520 °C for 15min in a semi-enclosed graphite-box containing about 0.3 g selenium particles by using a rapid thermal processing (RTP) furnace (MTI, OTF-1200X-RTP) under nitrogen flow about 80 sccm (the heating rate was about 8.6 °C/s).

**Solar cell fabrication.** The CZTSSe solar cells were prepared with a structure of glass/Mo/CZTSSe/CdS/i-ZnO/ITO/Ni/Al. Firstly, an approximately 40 nm thick CdS buffer layer was deposited onto the selenized CZTSSe film by a chemical bath deposition method. Then i-ZnO combined with indium tin oxide (ITO) window layers were deposited successively by radio frequency (RF) magnetron sputtering. Finally, the Ni-Al grid electrode was thermally through a metal shadow mask as a current collector. Notably, a 110 nm $MgF_2$ antireflection layer was evaporated onto the final cell to enhance the photocurrent. A solar cell device with an effective area of 0.18 cm$^2$ was separated by mechanical scribing, yielding 9 standard cells on the same substrate.

**Characterization.** Raman spectra were collected by a Raman spectrometer (LabRAM HR Evolution, HORIBA), using a 633 nm excitation laser with a power of 6 mW. The FT-IR spectra were collected by a Fourier Transform Infrared Spectrophotometer (TENSOR27, Bruker) and the KBr disk method. Solution $^{13}$C and $^{119}$Sn NMR spectra were recorded on a Bruker Avance III 400HD or Avance III 500 MHz spectrometer at 100.6 MHz or 186.5MHz. Spectra were obtained at room temperature with 10% percent of deuteroxide adding for locking of the main magnetic field. Each experiment of $^{119}$Sn, $^{13}$C used a 12.5 and 10 μs pulse width and a 1.0 and 2.0 s relaxation delay and collected 256 scans, respectively. A spectral window from 350 to -650 and 225 to -25 ppm was used to search for the signals of $^{119}$Sn and

$^{13}$C, respectively. δ $^{119}$Sn and $^{13}$C were referenced to the internal standard of the probe. The morphologies of the films and devices were obtained by a scanning electron microscope (S4800-SEM, Hitachi). Current density-voltage (J-V) characteristics of the solar cells were collected on Keithley 2400 Source Meter under AM 1.5G illumination (1000 W m$^{-2}$) from Zolix SS150A solar simulator. The light intensity of the solar simulator was calibrated by a standard monocrystalline silicon reference solar cell. The external quantum efficiency (EQE) curve was measured using an Enlitech QE-R test system, where a xenon lamp and a bromine tungsten lamp were used as the light sources and a certified Si and InGaAs diode was used as the reference detector. The C-V data was performed at 100 kHz and 14 mV alternating current (AC) excitation source with direct current (DC) bias ranging from 0.6 V to -1.0 V. Transient photovoltage spectra of the cell were obtained by a tunable nanosecond laser (Opotek, RADIANT 532 LD) with an ultralow light intensity of about 10 nJ·cm$^{-2}$ and recorded by a sub-nanosecond resolved digital oscilloscope (Tektronix DPO7104) with input impedances of 1 MΩ.

**Acknowledgements**

This work is financially supported by the National Natural Science Foundation of China (Nos. 51961165108, 51421002, 51972332, 51627803 , 51872321 and 11874402).

**Author contributions**

L. Guo, J. Shi, Y. Luo and Q. Meng conceived the idea, designed the experiments and did the data analysis. L. Guo and Y. Luo did the experiments and the data acquistion. Q. Yu, X. Xu and J. Zhou contributed to CZTSSe solar cell fabrications. B. Duan, J. Wu and Y. Li performed SEM, CV and TPV measurements. D. Li and H. Wu supported the CZTSSe fabricating, characterizations and discussions. L. Guo, J. Shi, Y. Luo and Q. Meng participated in writing the manuscript. All authors were involved in the discussions and approved the manuscript. L. Guo and J. Shi have contributed equally to this work.

**Competing interests**

The authors declare no competing interests.

**Additional information**

Supplementary information is available for this paper.